# Grain Boundary Defect Production during Successive Displacement Cascades on Nanocrystalline Tungsten Surfaces

Yang Zhang[1], Anus Manzoor[1], Predrag Krstic[2], Jaime Marian[3], Jason R. Trelewicz[1,2,4]*

[1]Department of Materials Science and Chemical Engineering, Stony Brook University, Stony Brook, NY 11794
[2]Insitute for Advanced Computational Science, Stony Brook University, Stony Brook, NY 11794
[3]Department of Materials Science and Engineering, University of California Los Angeles, Los Angeles, CA 11794
[4]Materials Science and Technology Division, Oak Ridge National Laboratory, Oak Ridge, TN 37831

*Corresponding Author: jason.trelewicz@stonybrook.edu

## Abstract

The interaction of radiation defects with grain boundaries (GBs) governs damage tolerance in refractory materials for extreme environments. Tungsten (W), a leading plasma-facing material for fusion, will be subjected to coupled ion and neutron irradiation that degrades both surface and bulk properties. In this study, molecular dynamics (MD) simulations are employed to examine defect evolution under successive 1 keV displacement cascades at a W surface in nano-bicrystals containing $\Sigma3\langle110\rangle\{112\}$ and $\Sigma5\langle100\rangle\{130\}$ symmetric tilt GBs. The free surface biases interstitial accumulation toward surface planes, reducing bulk interstitial populations, while vacancy saturation is driven by cascade overlap. When cascades indirectly interact with GBs, defect accumulation becomes strongly dependent on GB character. The higher energy $\Sigma5$ boundary acts as a more effective defect sink for interstitials relative to the coherent $\Sigma3$ boundary. This behavior arises from its larger interstitial segregation energy and enhanced strain field, which promote trapping via thermal migration and focused collision sequences. The deeper trap states of the $\Sigma5$ GB suppress interstitial emission and limit recovery, whereas the shallower traps and mobile crowdion configurations in $\Sigma3$ GBs enable dynamic defect recombination. These results highlight the critical role of GB structure and grain size in controlling radiation damage evolution in tungsten with behavior that applies broadly to refractory BCC metals.

## 1. Introduction

Understanding the role of interfaces in defect accumulation during irradiation represents a critical step forward in the design of damage tolerant nanomaterials[1-3]. Tailored microstructures incorporating grain and interphase boundaries provide energetically favorable sites for combating displacement damage due to irradiation[1, 4, 5] as well as cavity formation from gaseous impurities introduced by implantation or transmutation[6-12]. This mechanism for limiting damage accumulation plays a significant role in nanostructured materials due to their large interfacial area per unit volume[13], for which nanocrystalline body centered cubic (BCC) metals are of particular interest for extreme environment applications[14]. Tungsten (W) presents a unique opportunity to explore nanocrystalline structures for enhanced radiation tolerance as it has emerged as the leading plasma facing material for fusion devices[15-17] due to its combination of thermomechanical performance, resistance to sputtering, and chemical compatibility with tritium[5, 18, 19]. Indeed, various forms of nanocrystalline W have demonstrated an improved radiation resistance relative to their coarse-grained counterparts[20-25], suggesting that the high density of interfacial sinks helps to slow the rate at which surviving defects cluster into large loops and voids, which ultimately lead to swelling and embrittlement[26].

Point defects serve as the building blocks for these larger irradiation defects[27], and as such, understanding their generation and evolution – and subsequent interaction with engineered microstructural sinks – is crucial to tuning the radiation tolerance of a material. Computational methods, such as molecular dynamics (MD), have proven highly effective in investigating defect dynamics during displacement cascades[28]. For instance, Setyawan et al.[29] conducted MD simulations to study displacement cascades in W at various primary knock-on atom (PKA) energies and temperatures, resulting in the creation of a comprehensive database detailing surviving defects. Various types of defects are formed, including Frenkel pairs, vacancy and interstitial clusters, and dislocation loops[30-33]. Experimental characterization of such defects has relied heavily on in situ transmission electron microscopy (TEM) experiments, which have shown that 1/2 <111> vacancy and interstitial dislocation loops are the primary defects formed in thin W foils irradiated with $W^+$ ions to 0.01 dpa across a broad temperature range of 30-1073 K[32, 33]. Similarly, W exposed to 1 dpa neutron irradiation at 800°C exhibited dislocation loops and voids with average sizes of 5 nm and 31 nm, respectively, and the majority of the dislocation loops were of the 1/2 <111> type[30].

In polycrystalline metals especially with grain sizes in the nanocrystalline regime (i.e., generally less than 100 nm), displacement cascades interact with grain boundaries (GBs), significantly influencing the generation and recombination of point defects. Experimentally, interfacial characteristics that influence a GB's ability to act as efficient defect sinks have been explored using nanolaminate configurations with reports on the impact of interfacial structure[34, 35], proximity of the defects to the interface[36], and other intrinsic factors[37]. Of those intrinsic factors, the nature of the defects – whether a vacancy, self-interstitial atom (SIA), crowdion, extended defect cluster, or defect loop – also strongly influences the recombination behavior at an interface[8, 38, 39] with defect mobility exhibiting a strong dependence on temperature[40, 41]. For example, simulated displacement cascades near a $\Sigma 5$ twist grain boundary in silver observed the segregation of interstitial clusters to GBs via <110> crowdion motion[42]. Similarly, Park et al.[43] conducted displacement cascade simulations in single and polycrystalline W and reported that the recombination rate of vacancies and interstitials is faster in the absence of grain boundaries, while the number of surviving interstitials is twice as high in the polycrystalline microstructure due to attractive interactions between GBs and interstitials. In addition to the thermodynamic stability of point defects, GBs also influence the kinetics and mobility of these defects, with point defect mobility being either higher or lower compared to the bulk depending on the structure of the grain boundary[44].

Surfaces also serve as defect sinks and thereby impact the formation of primary damage induced by irradiation. Previous studies have indicated that the quantity of surviving vacancies and vacancy-type dislocation loops is greater in cascades influenced by free surfaces compared to those produced through PKAs initiated in the bulk of a crystals with periodic boundaries. For example, MD simulations by Aliaga et al.[45] revealed that <100> vacancy loops are significantly larger (up to 4 nm in size) and more prevalent in ion thin films as compared to bulk iron. A similar effect was reported in W where a large number of vacancies generated in the presence of a surface clustered into <100> and 1/2<111> vacancy loops near the surface[46]. The prevalence of vacancy loops can be attributed to the migration of interstitials towards free surfaces with their asymmetrical mobility resulting in vacancy clustering near these surfaces[45-47]. While the highlighted exemplary and other prior studies have demonstrated the influence of GBs and free surfaces on displacement cascades, to the authors' knowledge, investigations have not collectively considered the interactions of defects with GBs and surfaces, especially in the context of W-based materials.

In this study, we performed molecular dynamics (MD) simulations of impacting atoms on W free surfaces utilizing a nano-bicrystal model containing two distinct types of grain boundaries. Following the approach of Frolov et al.[48], we selected two symmetric tilt grain boundaries – Σ3<110>{112} and Σ5<100>{130} – representing low-energy grain boundary configurations with different tilt axes. An impact energy of 1 keV was chosen to induce cascades within the surface layers of the nano-bicrystals, facilitating cascade interactions with the free surface. The influence of the GBs was incorporated by adjusting the size of the nanograin to promote cascade interaction with the GB plane and immediately adjacent lattice planes. Our findings reveal that both GBs and surfaces significantly influence defect production with the presence of GBs near the surface region augmenting interstitial formation while vacancy production was largely unaffected.

## 2. Simulation Methods

### *2.1 Nano-bicrystal Models*

Bi-crystal models were employed to investigate the interaction of displacement cascades with grain boundaries in the presence of a free surface. Relative to typical bi-crystal models containing a single grain boundary for quantifying defect-grain boundary interactions during cascades [49-56], we required two grain boundaries to facilitate the introduction of a free surface while maintaining symmetry at the periodic boundaries normal to that surface. This configuration, referred to as a nano-bicrystal, is illustrated in Figure 1 with the location of the external impact atom shown relative to the free surface where the dark shaded region denotes the area over which this impact was confined. Two different columnar grain sizes of 12 and 6 nm, as shown in Figure 1(a,b) respectively, were considered where the grain size was defined as the distance between the two adjacent grain boundaries. The length of the grain boundary planes along the z-direction, and thus the effective thickness of the simulation cell, was held constant at 7 nm, which was confirmed to avoid cascade interactions with the fixed bottom layer of the simulation cell. Dimensions in the x- and y-directions were around 14 and 26 nm, respectively, and the initial structures contained approximately 170,000 atoms. Two different symmetric tilt grain boundary configurations were considered for both grain size models and included a $\Sigma3\langle110\rangle\{112\}$ and $\Sigma5\langle100\rangle\{130\}$ grain boundary. It is worth noting that the x and y dimensions are not simple multiples of grain size because they are determined by the d-spacing of the symmetric tilt GBs to maintain perfect lattice matching and periodic boundary conditions. Thus, four pristine structures were prepared for this study and denoted Σ3-12nm, Σ5-12nm, Σ3-6nm, and Σ5-6nm to capture the coincident site lattice (CSL) boundary type and effective grain size.

The nano-bicrystal models were first subjected to energy minimization and relaxation steps using molecular dynamics (MD) in the Large-scale Atomic/Molecular Massively Parallel Simulator (LAMMPS) platform [57]. The Tersoff bond-order interatomic potential derived by Juslin et al. [58] was employed in our simulations and has been shown to capture a realistic W-W interactions during nonequilibrium events such as collision cascades [59, 60]. To verify the accuracy of this this potential for our collision cascade simulations, we calculated the threshold displacement energy (TDE) along different crystallographic directions, defect energetics and morphologies via MD simulations. TDEs were 49 and 72 eV along the <100> and <110> directions, respectively, consistent with prior reports [61, 62] with the reference values of ~43eV along <100> and 70~85 eV along <110>, as shown in Supplement Table S2. Although the Tersoff potential is generally less accurate than Embedded-Atom Method (EAM) potentials in predicting static bulk properties near equilibrium, frequently underestimating vacancy formation energies and GB formation energies (Supplement Table S1), it offers distinct advantages in this domain, by explicitly incorporating the local coordination environment and angular dependence of bonds. it uniquely captures the most stable symmetry-broken <11ξ> self-interstitial configuration predicted by ab initio calculation from Ma et al. [63] (shown in Supplement Figure S1 and Figure S2) with consistent formation energy differences among <11ξ>, <100>, <110> dumbbell configurations (as shown in Supplement Table S1), and, critically, reproduces the reference Frenkel pair yield (2.1 ± 0.99 vs. 2.24 from Setyawan et al. [29]) and the reference displacement yield (31±5.77 vs 31~44 from Nordlund et al. [64]) for 1 keV cascades, as shown in Supplement

Table S3 and Figure S5. Energy minimization was conducted to achieve a final relative energy convergence of $10^{-12}$ with periodic boundary conditions applied in all directions. The structures were then relaxed through a simulated annealing step by heating to 650K at a rate of 0.1 K/ps using an isothermal-isobaric ensemble where it was held for 1 ns and subsequently cooled to 300K at the same rate. Finally, following the removal of the periodic boundary along z-direction, an additional relaxation was performed for 1 ns at 300K using a Langevin thermostat to achieve a zero-pressure condition along x- and y-directions.

*2.2 Collision Cascade Simulations*

Displacement cascades were initiated by accelerating a tungsten atom from 2 nm above the nano-bicrystal surface from a flat emission surface randomly along the 14 nm width of the model but constrained within a 3 nm region within the center of the grains. The velocity vector of these accelerated atoms was randomly oriented within a 7° cone relative to the surface normal to suppress channeling. This constrain impact region and cone of impact are highlighted in Figure 1. To localize the cascades within proximity to the surface, the incident tungsten atom was assigned a fixed kinetic energy of 1 keV. 1 keV PKA energy is relevant to low-energy DT or He plasma ions impacting tungsten surfaces in fusion reactors. Consistent with prior collision cascade simulations in BCC metals [29, 65-68], we neglected electronic energy losses, which are complex[69-71] and beyond the scope of this study. Cascade evolution employed the microcanonical NVE ensemble applied for a total of 10 ps for each cascade with a time step of 0.2 fs. Validation of 0.2 fs time step against a standard 0.1 fs time step for 10 keV cascades (more demanding than the 1 keV used here) showed no statistically significant differences in defect or temperature evolution (mean absolute percentage error of 8.7 % for defect counts and 0.3 % for temperature; Welch's t-test $p > 0.05$ at all time points), as shown in Supplement Figure S3 and S4.

Atoms occupying the lower three atomic planes were fixed during the cascade simulations using a freeze function. Applying a frozen layer at the bottom of the simulation cell is a standard procedure in surface irradiation simulations designed to prevent the artificial reflection of shockwaves from the bottom boundary and inhibit rigid shifting of the simulation box. The frozen region effectively mimics a semi-infinite bulk substrate, which better captures the physics of surface irradiation. Given the thickness of the simulation cell, the active region affected by the 1 keV cascade is located well above these frozen bottom planes. Furthermore, the kinetic energy from the PKA is efficiently dissipated prior to reaching these boundary layers, preventing any non-physical temperature buildup or reflection back into the active simulation region, thereby ensuring these fixed layers did not artificially influence the evolution of the cascades as well as the cascade-GB interaction. Following cascade evolution under fixed total energy and volume, the temperature of the system evolved to a temperature of approximately 325 K. At this point, the excess energy deposited in the system was extracted through an additional 4 ps relaxation at 300 K using a Langevin thermostat. Finally, the NVE ensemble was reapplied for 2 ps prior to initiating a subsequent cascade. This procedure was repeated for a total of 400 successive impact events. The maximum dpa at the end of the simulation is estimated in the range of 0.1 dpa to 1 dpa in the central section of the sample: ~0.2 dpa based on defect production, ~0.5 dpa based on the displacements or replacements of the atoms.

### *2.3 Bulk Defect Quantification under Cascade Induced Surface Reconstruction*

To produce quantitative defect trends as a function of the number of successive impacts, self-interstitial atoms and vacancies were indexed based on their occupancy of the Wigner–Seitz cells of a reference lattice. However, as shown in Figure 2a, the surface was restructured due to the impacting atoms colliding with the surface layers of the nano-bicrystal and involved adatoms from pile-up and surface accumulation concomitantly with the formation of depleted regions around the primary impact zone relative to the initial surface plane. These surface reconfigurations must be accounted for in the definition of our reference lattice as to avoid artificially inflating the number of interstitials and vacancies from the Wigner-Seitz occupancy analysis. We accomplished this by defining an extended reference lattice where the surface was "built-up" relative to the pristine structure through the addition of atoms that effectively extended the BCC lattice as shown in Figure 2b. These atoms formed a "cap" on the reference lattice relative to the pristine structure, which is used in mapping accumulated and depleted regions of the surface.

Atoms that accumulated dynamically above the original surface of the nano-bicrystal during each impact event conformed to the underlying BCC lattice. Thus, when compared with the extended reference lattice as part of the Wigner-Seitz occupancy analysis, these accumulated atoms were viewed as occupying BCC lattice sites and not quantified as interstitials in the defect analysis. Uniquely identifying these atoms, however, allowed for the number of atoms accumulated during the successive cascades to be quantified and distinguished from the total number of implanted atoms during impact. With the number of accumulated atoms all unoccupied sites above the damaged surface were then identified through comparison with the extended reference structure to distinguish the redistribution of surface atoms during the formation of a surface crater – denoted the "depleted region" – from bulk vacancies. This process effectively created an artificial "vacancy cap" on the damaged structure, as illustrated in Figure 2c, which was composed of the free space above the damaged surface of the impacted structure (relative to the extended reference structure) and the depleted region below the original surface of the pristine structure. With the former region fully occupied in the extended reference structure, we were able to designate this region as an extended vacancy cluster in the damaged structure and remove it from the defect analysis. This left only the portion of the vacancy cap occupying the depleted region, which was excluded from the bulk vacancy analysis but used in determining the number of atoms depleted from the pristine surface during successive collisions with sputtering factored into the net depletion trends. Analysis of defects in the grain boundary regions followed an analogous treatment with the grain boundary crystallography incorporated into the extended reference structure.

## 3. Results and Discussion

The structure of the nano-bicrystal models and impact energy of 1 keV were deliberately selected to allow for the free surface to influence cascade evolution and thus act as a sink for point defects and mobile defect clusters. We first consider this effect by mapping defect production and the transition to vacancy saturation within the primary damage zone using the larger 12 nm grain size structure where cascades evolve independently of the grain boundaries. These trends provide insights on surface reconstruction as an energy dissipation mechanism and provide a baseline for quantifying grain boundary effects in the presence of a free surface. Grain boundary defect production is then studied, capturing the mechanisms of grain boundary defect formation during the accumulation stages of the cascade as well as annihilation due to recombination and migration to the free surface driven by the successive cascade events. By comparing defect evolution behavior for the two different grain boundaries, we decouple mechanism-specific contributions to the cumulative grain boundary defect trends in the context of cascade driven events and diffusive processes biased by the grain boundary-dependent interstitial segregation energy. The GB interstitial segregation energy was computed as the energy difference between a self-interstitial atom in the perfect lattice and a SIA at various positions near the GB. The GB interstitial segregation energy showed a site dependency generally based on the distance to the GB, and the highest values (such as 6.5 eV for the Σ5 GB core) represent the strongest trapping site, as shown in **Figure 13**(c-d). The 6.5 eV of the core Σ5 GB SIA segregation energy is consistent with the first-principle calculations of 6.6 eV by Chai et al. [72], while core Σ3 GB SIA segregation energy of 3 eV is slightly higher than the first principle calculation of 1.87 eV by He et al. [73]. Therefore, even more pronounced differences in the SIA segregation behavior at the GBs would be expected with a more physically robust potential. It is also worthy of noting that many shallower meta-stable trapping sites exist further away from the center plane of the GB.

### *3.1 Lattice Defect Production Near a Free Surface During Single Cascades*

The cumulative number of defects during a single cascade on the (110) surface of the Σ3-12nm structure is shown in Figure 3a during the 10 ps period where the cascade evolved under constant system energy and volume. The damage region rapidly evolves as the recoil energy is transferred via atomic collisions to the lattice with a peak damage state produced at approximately 1.2 ps. From the snapshot of the cascade at its peak damage state illustrated in the inset, recoils evolved independently of the grain boundaries, and cascade evolution in these simulations is thus akin to the bombardment of a (110) surface of a perfect BCC W lattice. Magnified snapshots of the cascade's peak damage region (PDR), where most of the permanent displacements located, are shown in Figure 3(b-d) at the times indicated on the cumulative defect trends where atoms fully displaced from their parent lattice sites are indexed as larger red atoms while intermediate atomic displacements are shown as smaller atoms; all atoms displaced more than 0.5 Å are colored according to their atomic displacement. The snapshots indicate different phases during the cascade evolution as: (b) the early stage of the thermal spike phase, (c) 0.4 ps, peak stage of the thermal spike and the emission of the focused collision sequences (FCSs), which is a non-destructive sequence that translates the oscillation along preferable directions from the cascade core for several nanometers within a picosecond as if whole string of atoms experienced an elastic collision, and (d) 1.2 ps, peak stage of the thermal spike and the zipping back of the FCSs. As expected in the absence of subcascade formation at the lower impact energy of 1 keV[74], the many recoils lead to the formation of a spatially localized displaced core[27, 66], the PDR, which is outlined using a surface mesh as well as displacement sequence events emanating along favorable crystallographic directions well beyond this displaced core. Athermal recombination transpired up to ~4 ps with subsequent thermalization to the end of 10 ps energy conservative evolution of the cascade, and the final state prior to relaxation is shown in Figure 3e. Here, interstitials are indexed as large red atoms, vacancies as larger blue atoms relative to the smaller ones on the BCC W lattice. While the general morphological evolution of the cascade produced on the free surface was consistent with cascades produced in a perfect BCC lattice[27, 65, 66, 74-77], all surviving interstitials except 1 were located at the surface of the nano-bicrystal.

The displaced core and recoils extending beyond its periphery inevitably interacted with the free surface as seen in the inset of Figure 3a. Since surfaces act as effective sinks for point defects[78-80], the number of surviving defects will be augmented relative to bulk cascades as previously reported in Fe[81] and W[82]. For the impact on the pristine (110) surface of the Σ3-12nm structure shown in Figure 3, the number of surviving defects included 6 surface vacancies, 8 surface interstitials, and 3 lattice vacancies with no interstitials remaining in the lattice. Here, the surface interstitials were the atoms deposited on the pristine surface, and the surface vacancies were the sites under the pristine surface but connected to the vacuum, while the lattice defects were the defects below the restructured surface. The impacting atom was effectively implanted with two atoms sputtered from the surface. For a total of 10 distinct impacts on the pristine (110) surface, the average number of defects produced were 7.0 surface vacancies, 8.7 surface interstitials, 2.9 lattice vacancies, and <0.1 lattice interstitial. This biasing of point defects to the surface was also observed for single impacts on the (100) surface of the Σ5-12nm, which exhibited an average of 12.1 surface vacancies, 13.2 surface interstitials, 2.5 lattice vacancies, and <0.1 lattice interstitial. Inclusive of defects at the surface and within the bulk lattice, the total number of surviving defects produced by our surface impacts was 9.9 and 14.6 defects per impact for the (110) and (100) surface orientations, respectively. This increased number of defects for the (100) surface is attributed to three coupled mechanisms. First, the low kinetic energy of these 1 keV impacts prevents the cascade from developing fully isotopic, a significant fraction of the energy transfer is dominated by the first few collisions, which would be around the incident direction as shown in **Figure 3**c , and the TDE around <100> is lower than the TDE around <110> [61, 62]. Second, the proximity of the free surface forces the collision cascade into a hemispherical morphology. This broken spatial symmetry truncates the directional scattering that typically randomizes momentum in bulk cascades, thereby enhancing the dependency on the incident direction. Finally, from a thermodynamic perspective, the BCC {110} plane represents the most closely packed, lowest-energy surface. The higher-energy {100} surface is inherently less stable, presenting a lower thermodynamic barrier for surface restructuring, which naturally accommodates a higher yield of surface vacancies and adatoms during the thermal spike.

The single-impact defect population produced via a 1 keV surface cascade are generally greater than the reported values for bulk cascades simulations in W with comparable PKA energies[29, 65, 83, 84] and the values derived from experiment results[85-89]. Comparing with the database developed by Setyawan et al.[65], a 1 keV cascade in bulk tungsten produced 2.2 defects/cascade averaging over all directions. While the potential used in that study had a larger average TDE of 128 eV, comparing defect production at identical values of the PKA energy normalized by the TDE still did not account for the large disparity between the results obtained from our free surface simulations relative to bulk cascades. The presence of the free surface thus enhanced overall defect production with biased accumulation in the surface layers, whereas the number of interstitials surviving within the bulk was markedly reduced but with a comparable number of lattice vacancies. Similar behavior from 30 keV impacts on a tungsten surface was reported by Lee et al.[82] where a large surface concentration of interstitials was attributed to the surface sink effect and an energetic preference for the formation of adatoms. The resulting asymmetric interstitial mobility in the presence of the free surface suppressed point defect recombination, which produced many sub-surface vacancies and consistent with our results acquired under a lower PKA energy. Thus, relative to the reported trends for bulk cascades, defect production in the presence of the free surface is enhanced due to the biased accumulation of interstitials within the surface layers combined with suppressed defect recombination, which in turn reduced the number of interstitials within the bulk while having a limited impact on bulk vacancies.

### *3.2 Cumulative Defect Production and Vacancy Saturation*

Under cumulative cascade events, the characteristics of the damaged surface, combined with surviving lattice defects from prior cascades, ultimately affect the number of defects generated in a subsequent cascade. We note that the plateau in surviving defects shown in Figure 3 doesn't capture a fully thermalized state as long-range diffusion is limited on the timescales of the MD simulation. Therefore, rather than making broad categorical claims about grain size effects on radiation tolerance, we focus on the nature of the defects formed, their interactions with the two different grain boundaries, and a comparative analysis of the accumulation behavior. Figure 4a reveals the evolution of interstitials and vacancies during successive displacement cascades in the 12 nm nano-bicrystal containing the Σ3<110>{112} and Σ5<100>{130} CSL boundaries.

Consistent with the single-impact defect production described in the previous section, the number of accumulated lattice interstitials during successive displacement cascades remained extremely low, ~15 retained lattice interstitials after 400 successive impacts (red lines in Figure 4a). As shown in Figure 4b and c, these lattice interstitials are found outside of the vacancy-rich PDR located directly below the bombarded surface. These remote interstitials have been investigated by previous cascade simulation studies on a perfect BCC lattice[27, 29, 66, 75, 90-93] and attributed to two production mechanisms, displacement sequence events[77, 94-97] at low PKA energies or shockwave-induced mechanisms[27, 76, 98] at high PKA energies where subcascades may form, typically above 10 keV. The displacement sequence is described as a type of focusing mechanism[94, 95, 99-104] and therefore occurs along preferable crystallographic orientations, e.g., <111> directions in BCC. In our simulations, we observed the focused collision sequence as noted in Figure 3c. However, rather than creating a defect, the atoms immediately returned to their pristine positions following the oscillation in Figure 3d. Alternatively, replacement collision sequences (RCS) are initiated with an energy greater than the TDE and produce an interstitial at the end of the string. The presence of the free surface biased the production of lattice interstitials via the RCS mechanism to the periphery of the PDR while those near the surface contributed to the formation of ridges around the cascade region and collectively reduced the lattice interstitial production rate.

Examining the atomic structure of the interstitials identified through the Wigner-Seitz occupancy analysis in Figure 5a reveals several configurations: (i) a <100> dumbbell in Figure 5b, (ii) a mobile <11ξ> configuration in Figure 5c, and (iii) an interstitial cluster in Figure 5d. Among these configurations, the mobile <11ξ> configuration has a migration barrier of 0.040 eV [63], comparative to the <111> crowdion which has migration energies ranging from 2.6 meV[105-107] to 0.05 eV[108] and migration temperatures as low as 30K[105, 106, 109]. Under the present simulation conditions of 300 K and recognizing that SIA migration dominates the thermal defect annealing process up to 2050 K[110], the low migration barrier for the crowdion resulted in its spontaneous migration and subsequent formation of interstitial clusters as in Figure 4d. In contrast, vacancies migrate around 700 K[111, 112] with a migration energy of approximately 1.7 eV[107, 113, 114], and therefore vacancy clusters (e.g., divacancies formed from two monovacancies) were not observed as vacancies lacked sufficient mobility at this temperature. Given that interstitial mobility depends on their structure and configuration relative to the BCC lattice[39, 107], it will be these attributes that drive the accumulation of interstitials in grain boundaries, as well as intrinsic properties of the grain boundaries themselves.

Compared with the low interstitial production rate, the lattice vacancy concentration rapidly increased up to 50 cumulative impacts with the production rate from each impact matching the single impact cascade production from Figure 3. Following 50 impacts, a plateau was observed in the vacancy count with consistent fluctuations between the two structures (blue lines in Figure 4a). Defect saturation during successive cascades has been reported by Wu et al.[115] and Meyer et al.[116] and attributed to a reduction in the defect production rate after reaching a critical number of impacts. Gao et al.[117] also revealed that cascade overlap significantly reduced the defect production rate in BCC Fe. For a 2 keV cascade fully overlapped on the debris of a 2 keV cascade, the number of surviving defects were identical to the initial number in the cascade debris, i.e., the consecutive cascade didn't produce any additional surviving defects. They noted that pre-existing damage could shorten the RCSs and promote the recombination of Frenkel pairs within the cascade.

Vacancy saturation is therefore attributed to a balancing in the rate of defect production and recombination where the latter is enhanced in the previously damaged material. The number of surviving lattice vacancies was modeled under the assumption of reduced defect production driven by cascade overlap using an overlap probability ($P^n_{overlap}$). Assuming that the size of peak damage region was consistent throughout the simulation, the overlap probability of n^th impact falling into the previous PDRs can be calculated as:

$$P^n_{overlap} = 1 - P_{cas} = 1 - \left(1 - \frac{S_{PDR}}{S_{target}}\right)^{n-1} \tag{1}$$

where $P_{cas}$ is the probability that the $n^{th}$ cascade does not overlap on previous damaged zones, $S_{PDR}$ is the surface area of the PDR, and the $S_{target}$ is the bombardment target surface area for the initiation of cascades. In our simulations, $S_{PDR}$ is approximately 2 $nm^2$ (0.8 nm in radius) and $S_{target}$ is 52 $nm^2$. Since the PDR is small relative to the cascade initiation area, a binary approximation is applied with partially overlapped cascades ignored: a non-overlapping cascade produces a consistent number of vacancies ($N_{def}$) whereas no net new vacancies are produced in a fully overlapped cascade as discussed above. The number of surviving lattice vacancies after the $n^{th}$ impact, $N_v$, can be written as:

$$N_v = \sum_1^n \left(P_{cas} \times N_{def}\right) = \sum_1^n \left(\left(1 - \frac{S_{PDR}}{S_{target}}\right)^{n-1} \times N_{def}\right) \tag{2}$$

$$N_v = N_{def} \times \frac{1-(1-\mathrm{a})^n}{\mathrm{a}}, a = \frac{S_{PDR}}{S_{target}} \tag{3}$$

While binary approximation is an idealization (overlap may cause partial recombination or even new defect creation), it provides a minimal two-parameter description that captures the observed saturation behavior (as shown in **Figure 4**a or **Figure 6**a). Stochastic fluctuations are expected but do not alter the saturation trend. The number of lattice vacancies in the Σ3-12nm structure bombarded on a {110} surface was calculated using Eq. 2 with $N_{def}$ of 3 and shown in Figure 6a to capture the initial vacancy production regime and subsequent transition to defect saturation as the probability of a future cascade interaction with a prior PDR converges to unity. The Σ5-12nm structure with a {100} surface and $N_{def}$ of 2.5 exhibited identical saturation behavior consistent with the vacancy evolution trends in Figure 4a. The displacement defect saturation behavior was investigated by the electrical-resistivity measurement of the irradiated samples in early times, such as Birtcher et al. [118], the saturation of resistivity indicated the defect saturation. Recently, the study was facilitated by the transmission electron microscopy (TEM), the Doppler broadening positron annihilation spectroscopy (DB-PAS) measurement as well as the nano-indentation measurement, as in Wang et al. [119].

While defect accumulation is generally influenced by the cascade overlap effect, mechanisms occurring during the ballistic phase, such as sputtering (at a rate of 2.0 per impact) and implantation (at a rate of 1.0 per impact), maintained a consistent rate throughout the simulation as seen in the upper panel of Figure 6b. However, net surface accumulation and depletion, defined as the atoms accumulating on the surface ridges less those from implantation, and the atoms displaced from the surface valley less those from sputtering, respectively, are shown in the lower panel of Figure 6b to exhibit a rate shift consistent with the transition to cascade overlap, and finally a plateau with stochastic fluctuations. This rate shifts are governed by a transition between two distinct physical regimes. In the initial stages, the collision cascades restructure the pristine surface, driving direct lateral mass transfer from the surface valley to the adjacent ridges. As the cascades begin to fully overlap the target region, the subsurface lattice becomes highly defected and 'looser' due to a dense concentration of stable vacancies, and the periphery becomes denser due to the accumulation of interstitials. Incoming kinetic energy is increasingly consumed by localized rearrangement and recombination within the volume of PDR, slowing the outward displacement of mass, while high density of interstitials at the PDR periphery promotes the interstitial migration to the sinks, such as the surface (plus GBs in the later settings).

Ultimately, the system reaches a topological limit: as the relative altitude between the deepening surface valley and the growing ridges becomes too high, the steep crater walls physically block lateral mass transfer. Displaced atoms can no longer clear the ridges and are instead re-deposited within the valley. With lateral transfer shut down, sputtering becomes the sole active mechanism for surface depletion (considering vacancy saturation in the subsurface region), causing the net depletion rate to become identical to the sputtering rate. Simultaneously, although direct valley-to-ridge mass transfer is blocked, the cascades continue to introduce extra mass via implantation. These implanted atoms are pushed to the periphery of the PDR as interstitials. Driven by the strong thermodynamic sink of the free surface, these excess interstitials are continuously 'pulled' from the PDR periphery up to the surface. Therefore, at the later phase of surface morphology evolution, surface accumulation is maintained not by lateral mass transfer above the surface, but by the steady-state migration of excess interstitial atoms under the surface, perfectly balancing the disparities between lattice vacancy generation, implantation, and sputtering. The comparison between different sized structures also indicated the under-surface interstitial migration led to the surface accumulation. With the 6nm grain size settings, interstitials were blocked by the GBs, which also limited the growth of the surface ridge both the mass and the size. At the end of 400 impacts, 14.3% of 672 surface accumulations extended beyond the ± 3.5 nm on the surface of Σ5-12nm structure while only 8.2% of 563 surface accumulations extended beyond the ± 3.5 nm (where the GB presented) on the surface of Σ5-6nm structure. Similarly for Σ3 structures, 12.1% out of 618 on the surface of Σ3-12nm structure vs 7.6% out of 615 on the surface of Σ3-6nm structure reached beyond Σ3-6nm GBs at the ± 3.1 nm.

*3.3 Biased Defect Production in Grain Boundaries*

The reduction in the nano-bicrystal grain size to 6 nm positioned the GBs closer to the bombardment area such that FCS events will interact with the GBs without the GBs being directly damaged by the impact atoms. Cumulative defect evolution is shown in Figure 7a and reveals that while the vacancy trends were comparable, interstitial accumulation varied with GB type. Specifically, the Σ5-6nm GBs experienced an increased rate of accumulation relative to the Σ3-6nm GBs, resulting in the former being populated by a higher density of interstitial defects as evidenced in Figures 7(b,c). Vacancies accumulated in the PDR concentrated within the grain interiors, which when combined with the GB structure remaining intact following 400 cascades, indicates that the impact atoms did not directly impinge on the GBs. Similar behavior was reported by Pérez-Pérez and Smith[55] on cascade evolution near GBs in Fe, where the GB captured some interstitials via the RCS mechanism rather than being directly reconstructed through a cascade overlap mechanisms. The remaining analysis will therefore focus on decoupling the role of GB properties from cascade dynamics on interstitial accumulation in the two different GB structures.

It has been widely reported that the accumulation of interstitials or vacancies in the GB inevitably alters the GB structure resulting the evolution to new microstates[48, 120-124] and/or migration of the GB[125-127]. To systematically determine defect distributions in the damaged structures, the modified WS analysis described in the method section was applied to delineate defects in the GBs from lattice defects by explicitly indexing atoms occupying GB sites in the nano-bicrystal. Cumulative lattice defect trends are shown in Figure 8a with the vacancies matching the total defect production from Figure 7a, consistent with the observation that vacancies remained within the PDR in the bombarded grain, whereas the lattice interstitial population was markedly reduced relative to the total number of interstitials. Interestingly, the lattice interstitial trends for the 6 nm nano-bicrystal closely resembled the cumulative interstitial trends for the 12 nm structures, indicating that GBs biased the defect production process. Shown in Figure 8b are trends for GB interstitial accumulation, where enhanced GB defect production was evidenced in both structures with the Σ5<100>{130}GB exhibiting a rapid increase at ~100 impacts, which was delayed to ~200 impacts in the Σ3<110>{112} GB and resulted in the Σ5 GB collecting more interstitials after 400 cumulative impacts. GB dependent accumulation behavior in the lattice has been previously reported by Esfandiarpour et al.[52] and attributed to the stronger interstitial absorbency enhancing vacancy production in the lattice. Zhang et al.[53] reported variations in the lattice vacancy production rate from the cascades on different GB structures due to detailed interactions between defects and the GB. In the section that follows, defect production mechanisms deriving from cascade-GB interactions are further elaborated on in the context of the shockwave model for collision cascade evolution[76] collectively with cascade-defect interactions described by Wang et al.[128]

### *3.4 Grain Boundary Defect Evolution During Successive Cascades*

Examination of all GB interstitial creation events identified two primary mechanisms for GB interstitial production: (1) the formation of an RCS due to the trapping of an FCS by the GB during the ballistic phase of cascade evolution, and (2) thermal interstitial migration following the recombination phase. The first mechanism is illustrated in Figure 9 for cascades generated just adjacent to both GB types produced FCS events (evident in the peak damage state in Figures 9a and 9c) that impinged on the GB resulting. Interaction of the FCS with the GB transformed it into a RCS that ultimately produces an interstitial in the GB plane, as indicated by the atomic displacement vectors in Figures 9b and 9d. According to the shockwave model, when the shockwave front with a higher atomic density encounters an obstacle, its progression halts. The trailing part of the shockwave then compresses the leading part, leading to the formation of an RCS. These RCSs generate interstitials near the obstacle and vacancies near the cascade core. In our simulations, due to the low-energy PKAs, shockwaves appeared as focusing chains, and the transmission of the FCS was obstructed by the GB. This obstruction led to the formation of RCSs between the PDR and the GB, resulting in interstitials at the GB and vacancies in the PDR. The interstitials were then trapped at the GB by the segregation energy (as shown in Figure 13), preventing their emission without additional excitations. The rate of this RCS formation induced GB interstitial production was identical for both GBs (solid red line in Figure 13), though the total interstitial production rates differed.

To further clarify this mechanism, a controlled collision sequence was generated by accelerating an atom 2 nm away from the GB plane along the <111> direction at toward the GB with energies of 1-70 eV. A time sequence of its evolution is shown in Figure 10a for the Σ5<100>{130} GB with the final sequence magnified in Figure 10b. The results indicated that the threshold displacement energy required to create an interstitial decreased from 64 eV in a perfect lattice to 55 eV in the presence of the Σ5 GB. However, the controlled collision sequence revealed a different behavior for FCS trapping as compared with the bombardment simulations. When the GB trapped the focusing chain, a vacancy was created adjacent to the GB. In contrast, vacancies were created in the PDR during the collision cascade simulations, as described above. This suggests that the cascade influences focused collision trapping behavior, likely by compressing the trailing part of the FCS or by energizing the lattice, resulting in a lower TDE. Nevertheless, when GB interstitials are formed, the resulting configurations are influenced by the GB structure, as demonstrated in Figures 10c and 10d. In the $\Sigma3$ GB, the interstitial existed in a mobile configuration, while in the $\Sigma5$ GB, the interstitial may be trapped in a stable pentagonal cage. This difference suggests that the trapping strength for GB interstitials varies between the two GB types, leading to different rates of interstitial emission from the GB.

The second mechanism for GB interstitial formation, thermal interstitial migration, is illustrated in Figure 11 for the two different GB configurations and distinct types of defects that formed in the lattice during cascades. In Figures 11(a,b), a crowdion cluster migrated to the $\Sigma5$ GB along the <111> direction and was subsequently absorbed via local restructuring of the GB. In contrast, as shown in Figures 11(c,d), a <111> crowdion near the $\Sigma3$ GB located at a similar distance away from the GB plane transformed via a rotation into a more stable <11ξ> dumbbell configuration rather than migrating to the GB. As noted by Ma et al.[63] and Ventelon et al.[129], the symmetry-broken self-interstitial configuration in W, known as the <11ξ> dumbbell, shares key characteristics with the <111> crowdion configuration observed in the present study (cf. Figure 5c). The migration of the <11ξ> dumbbell begins with its transformation into a <111> crowdion. Therefore, the lattice structure near the GB seemingly influenced the interstitial migration behavior.

Considering both mechanisms for GB interstitial formation collectively, the migration-based GB interstitial formation process occurred more frequently in the nano-bicrystal containing the $\Sigma5$ GBs as compared with the $\Sigma3$ GB structures. In addition to the inherent crystallographic orientation differences for each GB and the surrounding lattice, varying interstitial segregation energies also influence defect accumulation. Interstitial segregation energy is defined as the energy change when moving a lattice interstitial into a GB atomic site. Uberuaga et al.[44] demonstrated that the flux of defects to a GB, relative to the flux of defects to an ideal sink, affects the "sink efficiency," which is influenced by both the defect migration barrier and the defect binding energy at the GB (analogous to the segregation energy discussed in the present study). A GB with higher sink efficiency behaves more like a perfect defect sink, efficiently removing defects from the lattice over time. While the short simulation times in MD do not permit the final defect state to be captured, it is expected that the sink efficiency of the GB would influence the rate of defect accumulation over the same period of time given differences in the interstitial segregation energies.

The number of surviving GB defects will also depend on their propensity for recombination during consecutive cascade events. Two such examples are demonstrated in Figure 12 and derive from the emission of GB interstitials following excitation due to a subsequent cascade. This process is illustrated in Figure 12(a-d) where the highlighted GB interstitial (12a) becomes perturbed from its sessile state within the first picosecond during cascade evolution (12b,c). Upon thermalization, the interstitial is ejected from the GB and recombines with a nearby lattice vacancy (12d). Alternatively, if the GB interstitial was in proximity to the surface, it would have an affinity to migrate toward the surface and contribute to surface accumulation, as demonstrated in Figures 12(e-h). While in both cases the interstitials were initially sessile dumbbells at the GB, the shockwave from the cascade drove the interstitial to reconfigure into a crowdion configuration with an intrinsically higher mobility. These crowdions then migrated into configurations with lower energy, allowing for eventual recombination with a nearby vacancy or annihilation at the surface. GB interstitials have been shown to recombine with nearby lattice vacancies along preferential migration directions through a replacement process with significantly smaller energy barriers[50]. However, given the differences in occurrence rates, the presence of nearby vacancies cannot be the sole factor responsible for the interstitial emission observed in the present study. Our findings indicate that both the proximity of interstitial sinks (such as the surface and vacancies) and the energy required to activate the GB interstitial contribute to the emission process, which were intuitively influenced by the GB configuration.

*3.5 Mechanistic Contributions Driven by Segregation Energy Landscapes*

To understand the role of local GB effects on defect evolution, the different contributions to GB defect production during the cumulative cascades were delineated in Figures 13a and 13b for the Σ3-6nm and Σ5-6nm nano-bicrystal structures, respectively. Following the discussions in prior sections, the mechanisms of GB defect production included: (i) RCS formation due to GB trapping of an FCS, (ii) interstitial thermal migration to the GB, (iii) GB interstitial migration to the surface, and (iv) GB interstitial recombination with nearby lattice vacancies. Considering first the contributions from the two GB interstitial emission processes of recombination and migration to the surface, both played a larger role in cumulative defect production for the Σ3 GB in Figure 13a relative to the Σ5 GB in Figure 13b, which didn't exhibit a large number of emission events until the GB was highly saturated with defects. From a configurational perspective, interstitials in the Σ5 GB were less likely to be energetically activated compared due to their sessile configurations while those in the to those in the Σ3 GB were in a more mobile crowdion-like configuration, thereby promoting emission during neighboring cascade events with the net effect of reducing the GB defect accumulation.

GB defect production for cascades initiated within neighboring grains (i.e., with the PDR removed from the GB plane) occurred via the previously described RCS formation and thermal migration mechanisms. The contribution from RCS to GB interstitial accumulation was independent of GB type as both GBs effectively acted as FCS trapping sites. Conversely, interstitial migration to the Σ5 GB in Figure 13b contributed more strongly to GB defect production relative to the Σ3 GB in Figure 13a. This vastly different migration behavior can be effectively explained through GB interstitial segregation energy landscapes, as shown in Figures 13c and 13d. for the Σ3-6nm and Σ5-6nm nano-bicrystal structures, respectively, superimposed over the local atomic structure from the GB into the grain interior. The segregation energy profiles both exhibited a smooth monotonic decrease from the grain interior into the GB plane with the transition region attributed to the lattice strain imposed by the GB as indicated by the region of local compression (reduced atomic volumes) directly adjacent to the GB. However, the segregation energy for the Σ5 GB was more than double the value for the Σ3 GB (6.5 eV vs. 3 eV, respectively), indicative of a higher sink efficiency. When combined with the accentuated lattice strain field, which will drive defect migration, the Σ5 GB will accumulate interstitials more effectively as reflected in the accentuated migration component of GB defect production. Once an interstitial occupies a GB site in the Σ5 GB, the larger segregation energy will promote retention of those defects, reducing emission and recovery through long range recombination and/or surface migration.

The general sink behavior of GBs has been reported throughout the literature[130-135], often highlighting the effects of GB structure and energy landscape on defect accumulation. Our findings add to this understanding by correlating the interstitial segregation energy and local lattice strain to the mechanisms responsible for interstitial production in the GB. These findings are consistent with other computational studies[73, 136] that have demonstrated higher-energy, structurally complex boundaries, such as the Σ5 studied here, act as more efficient defect sinks as compared with highly coherent boundaries like the Σ3 due to their expanded free volume and distinct interfacial energy landscapes.

Specifically, He et. al.[73] revealed a negative linear correlation between the excess volume of the GB and the interstitial segregation energy among 8 different symmetric tilt GBs. Li et. al.[136] calculated the annihilation energy barriers along different paths between vacancy and interstitial loaded tungsten interfaces, showing that the energy required for this type of recombination varied between paths. For a vacancy near a Σ5<100>{210} GB and an interstitial at the core of the GB, the energy barrier was 0.35 eV whereas it was only 0.04 eV along the preferential path for an interstitial on the (110) surface to a vacancy beneath the surface. Although this Σ5 configuration was slightly different from the Σ5 examined herein, both GB configurations are revealed to not only traps defects more rapidly, but also suppress subsequent interstitial emission and localized healing, which contrasts sharply with the shallower trap depths and highly mobile crowdion-like configurations that drive interstitial emission and dynamic recovery in Σ3 boundaries. Taken collectively, this behavior is consistent with the experimental observation of larger void denuded zones for non-Σ3 GBs relative to the highly coherent Σ3<110> tilt boundaries (coherent twin)[34] as a manifestation of the higher sink efficiency for higher-energy, structurally complex GBs.

## 4. Conclusions

MD simulations of cumulative cascades on W surfaces in the presence of two different CSL GBs – Σ3<110>{112} and Σ5<100>{130} – revealed two key sources of biased defect production and evolution: (1) the free surface enhancing overall defect production via interstitial accumulation in the surface atomic planes, and (2) cascade – GB interactions leading to biased interstitial accumulation in the GBs. In the larger 12 nm nano-bicrystals where the grain boundaries were sufficiently removed from the cascade damage region, the defect production rate in the PDR was governed by the defect formation energy, owing to the limited long-range defect generation and increased in-cascade recombination. Since the cascades on a previously damaged region promoted the recombination of the defects within the PDR during the energy deposition and dissipation phases successive impacts on the pre-damaged region significantly lowered the average rate of new defect production. After achieving a critical fluence, the number of defects in the PDR plateaued, as the overlap probability approached unity. Dynamic crowdion migration facilitated the creation of interstitials outside the core of the cascade but without an increase in the number of lattice vacancies in the PDR. Localized mass transport resulted in the reconstruction of the W surface, which was accounted for in the defect quantification through a modified WS analysis method.

By reducing the grain size of the nano-bicrystal structure to 6 nm such that the periphery of the cascades indirectly interacted with the GBs, the activation of additional defect production mechanisms led to the accumulation of interstitials within the GBs. The evolution of these defects was understood in the context their formation mechanisms including thermal migration of point defects and RCS formation due to GB trapping of an FCS. Balancing the creation of GB interstitials was their emission and recombination with lattice vacancies as well as migration to the free surface. The Σ5 GB exhibited delayed emission and enhanced interstitial accumulation through thermal migration, collectively explained by larger lattice strains driving this migration with the increased defect segregation energy (or binding energy) promoting the retention of defects relative to the Σ3 GB. Our findings therefore revealed that, relative to bulk cascades, distinct defect production behavior results from the introduction of grain boundaries in the presence of a free surface due to biased GB defect accumulation driven by the intrinsic properties of the GB and its impact on local lattice distortions as contributors to the effective sink efficiency.

**Acknowledgements**

This work was supported by the U.S. Department of Energy Scientific Discovery through Advanced Computing (SciDAC) Program under Award DE-SC0024401. Simulations were conducted using the Theory and Computation facility of the Center for Functional Nanomaterials (CFN), which is a U.S. Department of Energy Office of Science User Facility, at Brookhaven National Laboratory under Contract No. DE-SC0012704. Additionally, this research used resources of the Oak Ridge Leadership Computing Facility at the Oak Ridge National Laboratory, which is supported by the Office of Science of the U.S. Department of Energy under Contract No. DE-AC05-00OR22725

**Figures**

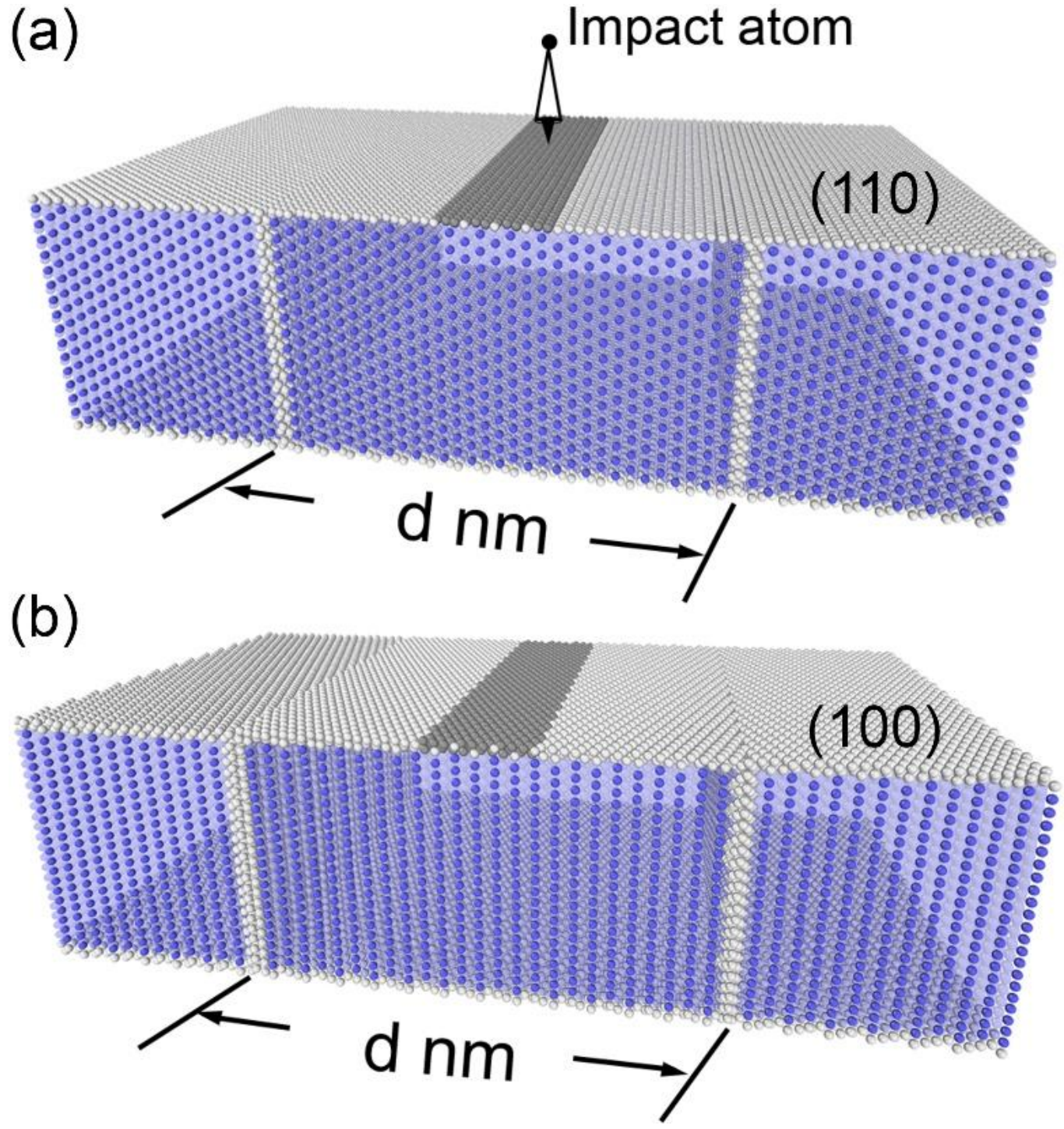

**Figure 1.** Tungsten nano-bicrystal structures containing (a) the Σ3<110>{112} CSL GB and (b) the Σ5<100>{130} CSL GB. The crystallographic orientation of both surfaces is shown with the region over which impact atoms were introduced shaded dark grey. Grain boundary and surface atoms are colored grey and atoms composing the BCC tungsten lattice blue. The grain size, d, varied between 6 and 12 nm to manipulate cascade interactions with the grain boundaries.

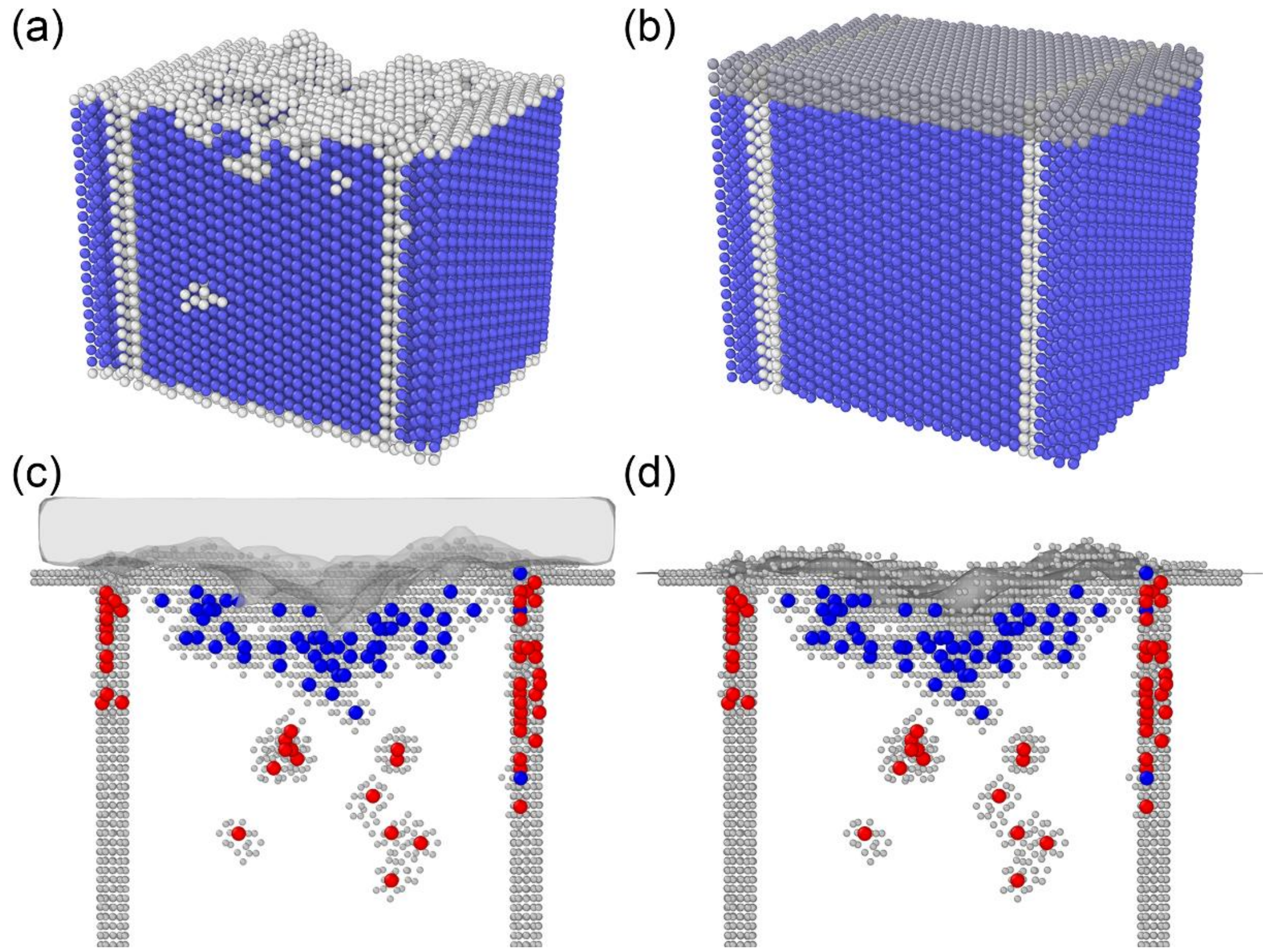

**Figure 2.** Defect quantification employing an extended reference lattice to account for surface restructuring during the successive impact events. (a) A representative damaged structure after 400 cumulative impacts demonstrating both surface accumulation and the formation of a crater in the primary damage zone. (b) Extended reference lattice where atoms were added to create a 2 nm-thick atomistic cap colored as dark grey. (c) Identification of the surface "vacancy cap", which is shown removed in (d) with bulk interstitials colored red and vacancies blue.

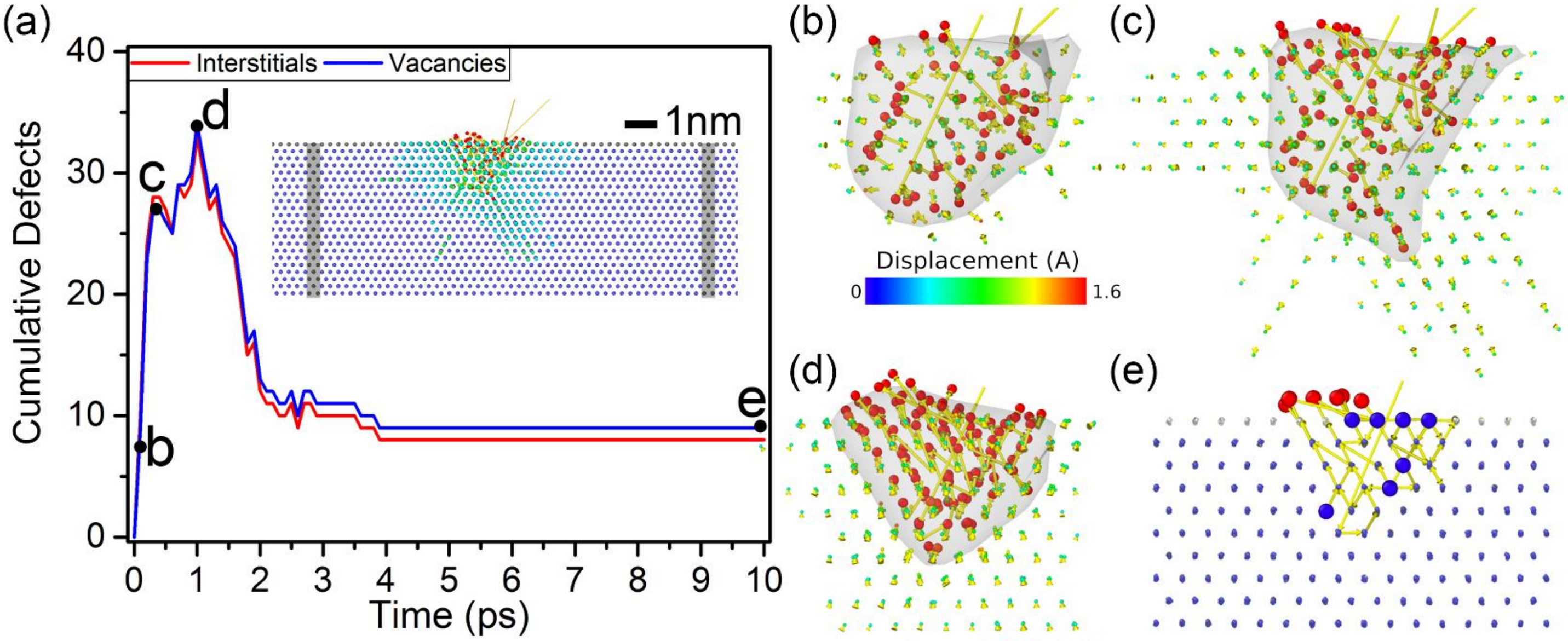

**Figure 3.** The time evolution of a single collision cascade induced on the (110) surface of the 12 nm grain size tungsten nano-bicrystal containing Σ3<110>{112} grain boundaries. (a) Total number of interstitials and vacancies as a function of time from the initial cascade event with a snapshot of the cascade shown at point c in the inset. Magnified snapshots of the PDR at simulation times of (b) 0.2 ps, early stages of the thermal spike phase, (c) 0.4 ps, emission of the FCSs, and (d) 1.2 ps, recovery of the FCSs, with atoms colored according to the magnitude of their displacement for values > 0.5 Å. Atoms fully displaced from their lattice sites are indexed as large spheres while atoms exhibiting intermediate displacements are shown as smaller spheres. (e) Final defect state at 10 ps after impact. The larger atoms are defects with interstitials in red and vacancies in blue.

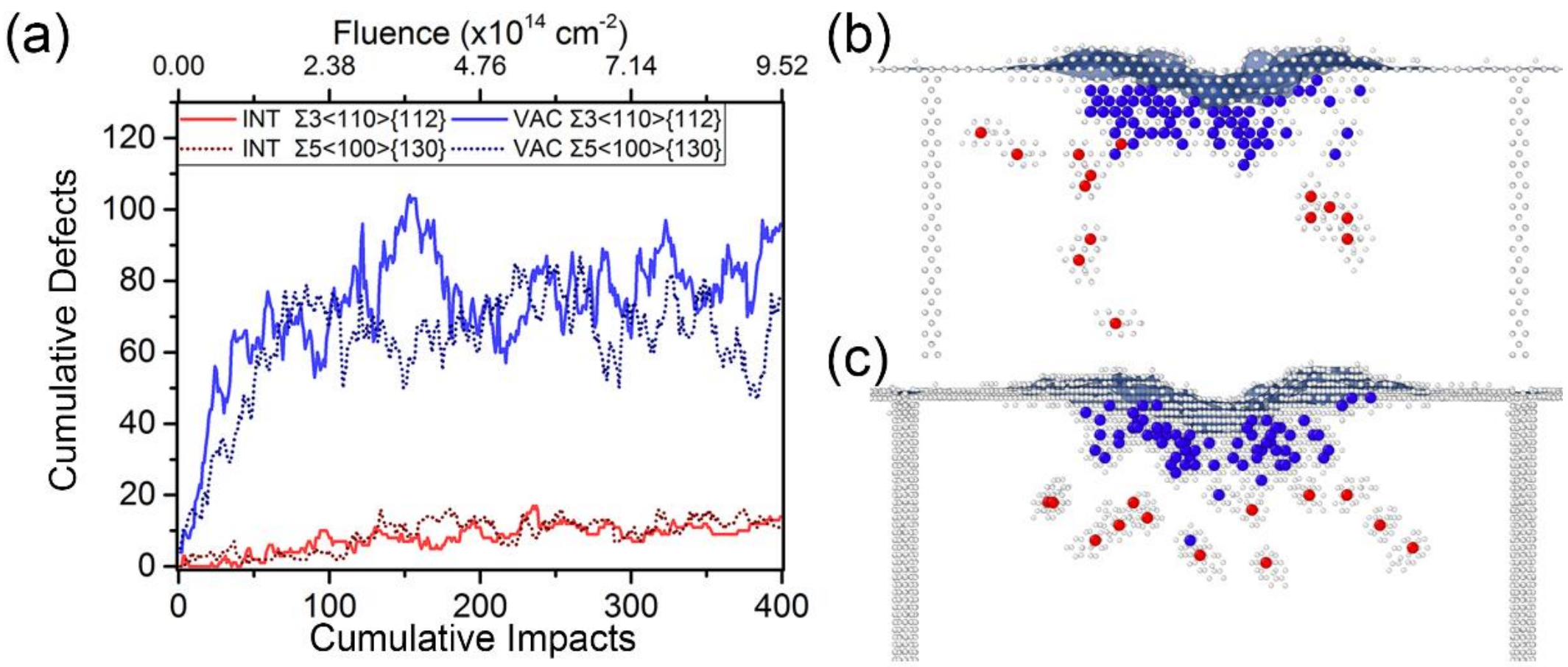

**Figure 4.** (a) Evolution of interstitials and vacancies during successive displacement cascades in the 12 nm nano-bicrystal containing Σ3<110>{112} (solid lines) and Σ5<100>{130} (dashed lines) GBs. Distributions of vacancies (blue atoms) and interstitials (red atoms) in the (b) Σ3<110>{112} and (c) Σ5<100>{130} structures; GB and surface atoms are colored grey and lattice atoms removed for clarity.

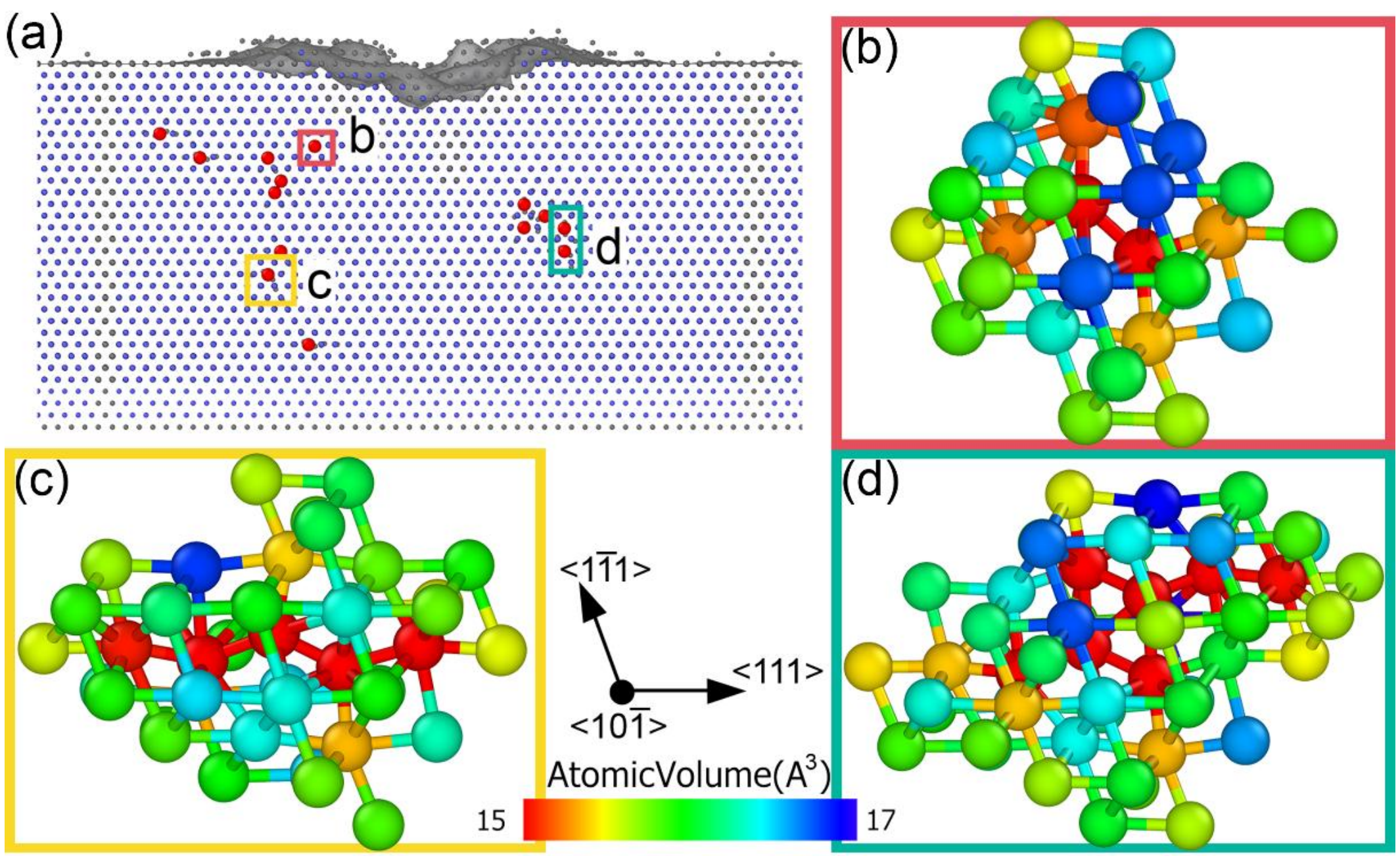

**Figure 5.** (a) Interstitial distribution in the 12 nm tungsten nano-bicrystal containing Σ3<110>{112} grain boundaries with BCC atoms colored blue, grain boundary atoms grey, and interstitials red. Atomic configurations of the interstitials identified in (a) from the Wigner-Seitz analysis were marked with colored boxes: (b) a <110> dumbbell configuration, (c) a mobile <11ξ> configuration, and (d) an extended interstitial cluster. In (b-d), the atoms are colored by the atomic volume based on a Voronoi construction.

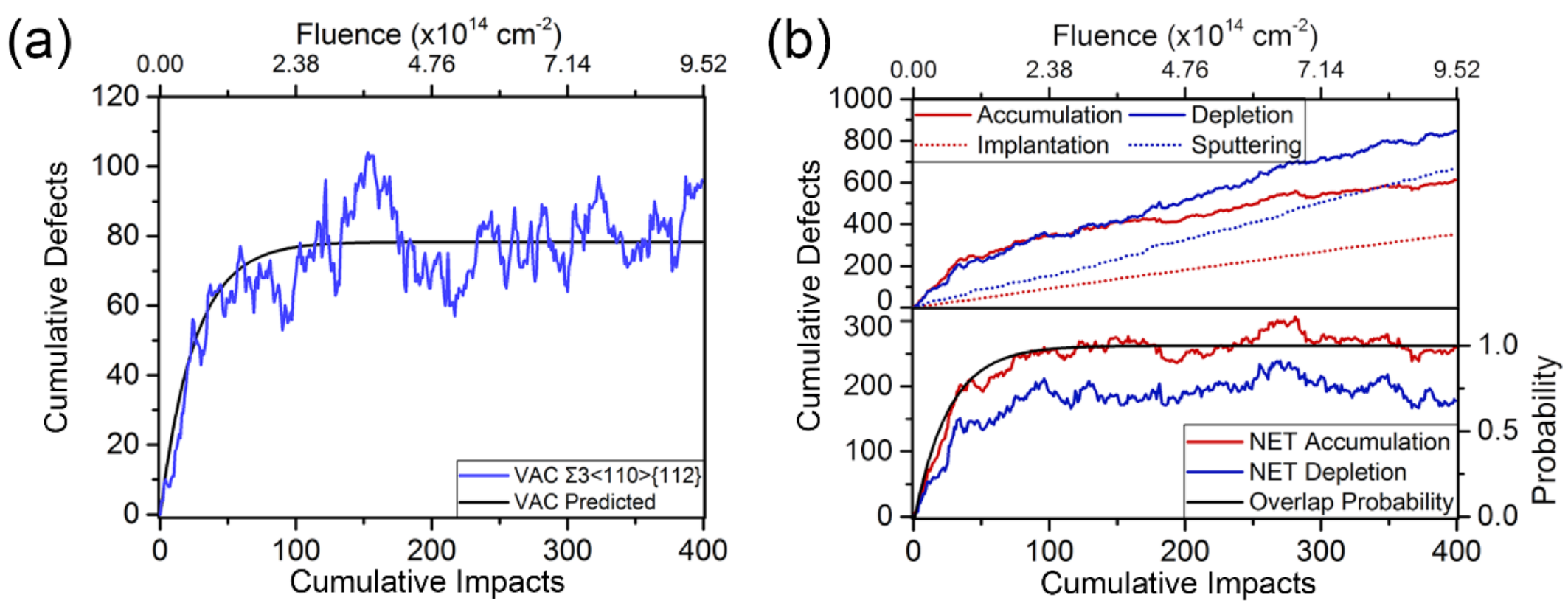

**Figure 6.** (a) Lattice vacancy accumulation in the 12 nm tungsten nano-bicrystal containing Σ3<110>{112} GBs with the predicted vacancy accumulation trend from the cascade overlap probability model. (b) Net accumulation and depletion trends in the lower panel as determined from the surface restructuring in the upper panel. The net accumulation and depletion trends follow the overlap probability, which is also plotted in the lower panel.

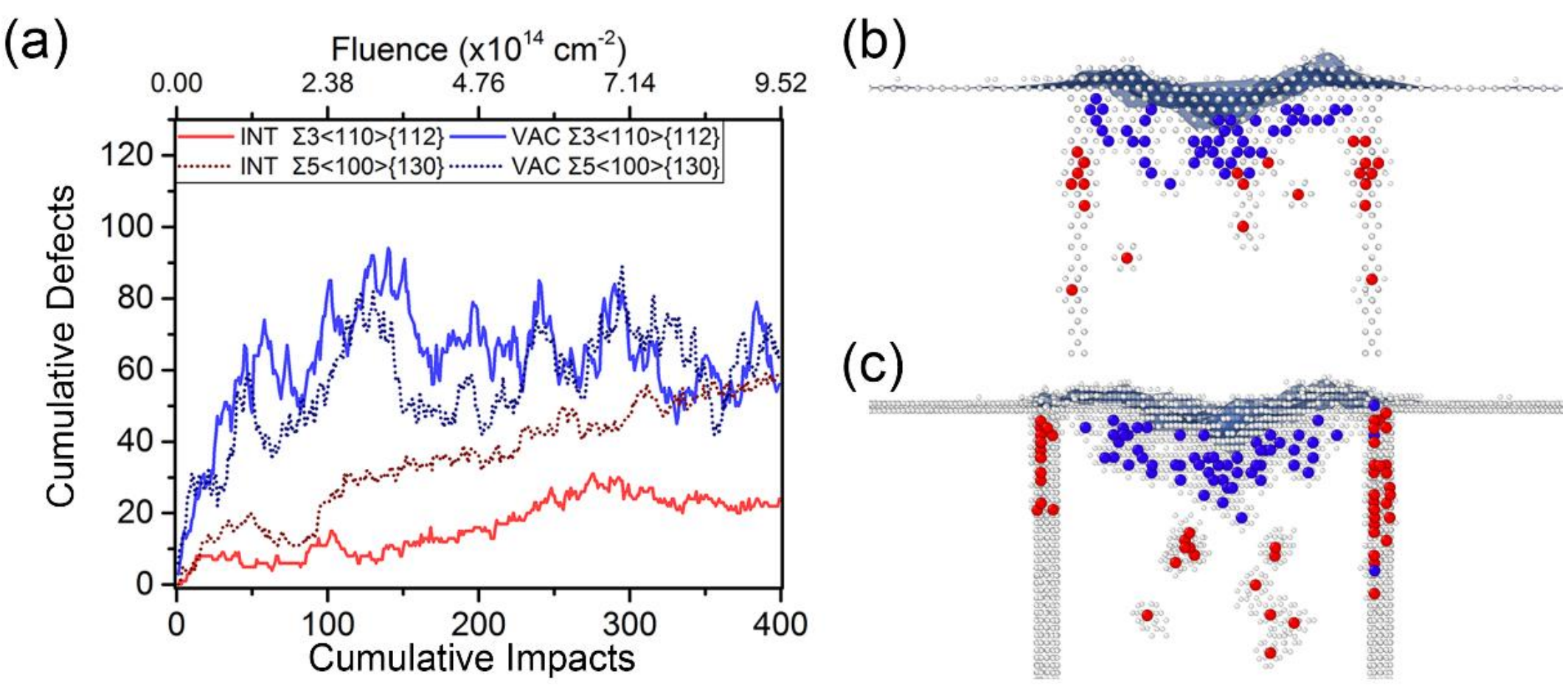

**Figure 7.** (a) Accumulation of interstitials and vacancies during successive cascades in the 6 nm nano-bicrystal containing Σ3<110>{112} (solid lines) and Σ5<100>{130} (dashed lines) GBs. Distributions of vacancies (blue atoms) and interstitials (red atoms) in the (b) Σ3<110>{112} and (c) Σ5<100>{130} structures; GB and surface atoms are colored grey and lattice atoms removed for clarity.

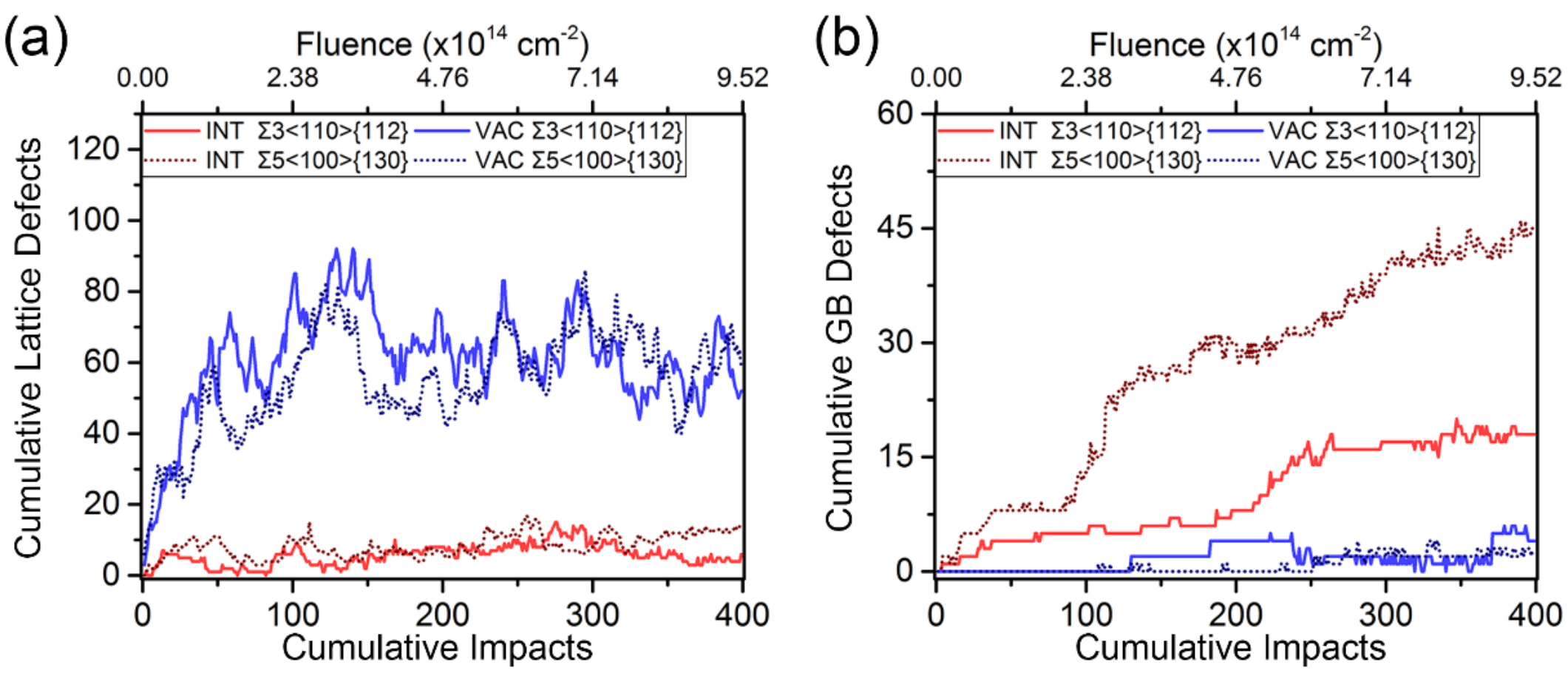

**Figure 8.** Distribution of point defects within (a) the BCC grain interior (lattice) and (b) the GBs during successive collision cascades in the 6 nm nano-bicrystal. Accumulation of interstitials is largely biased to the grain boundaries whereas vacancy formation is concentrated within the tungsten lattice.

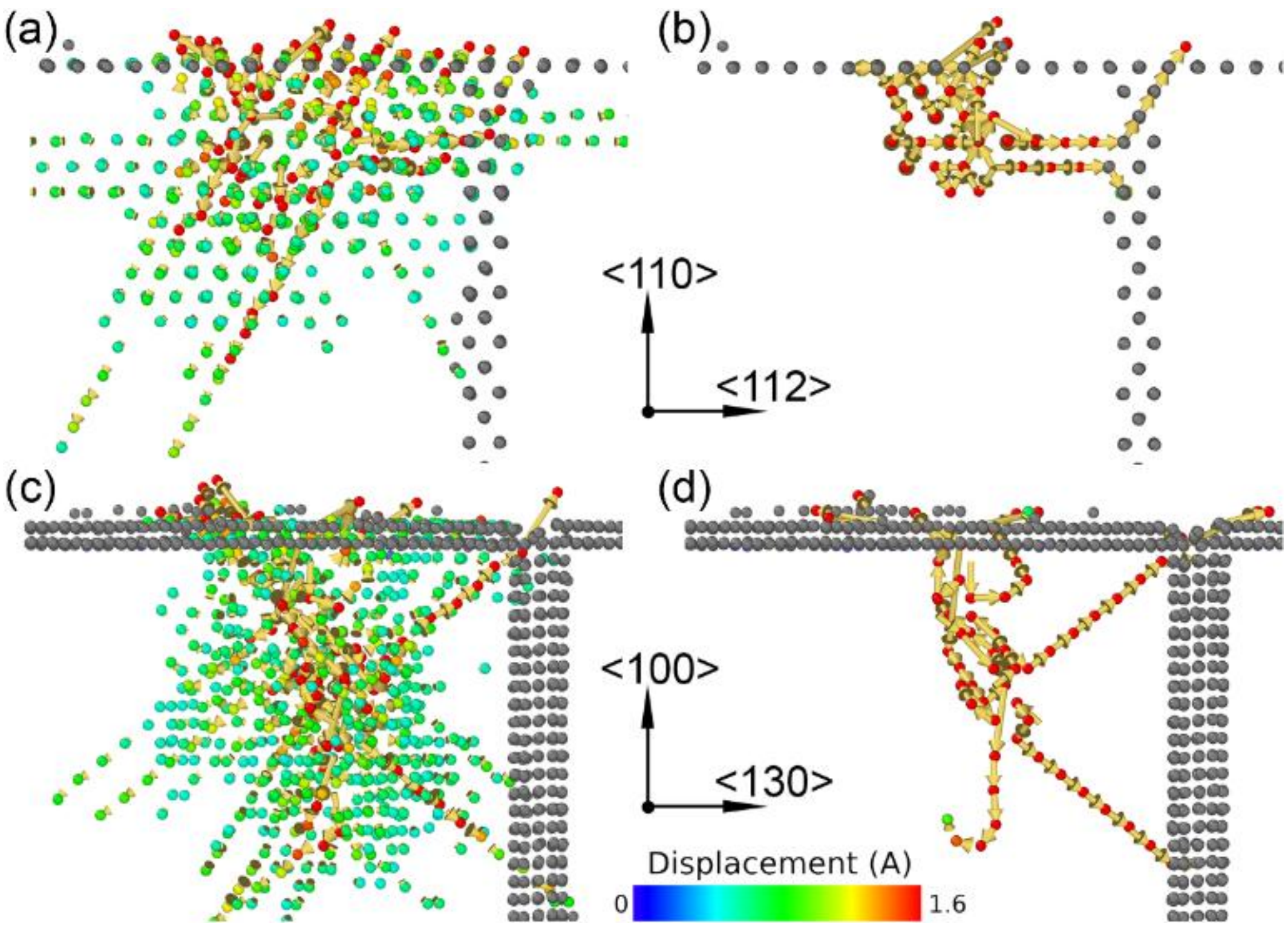

**Figure 9.** Snapshot of a collision cascade in the 6 nm nano-bicrystal structures. (a,b) The Σ3<110>{112} GB at the peak damage state and after 10 ps . (c,d) The Σ5<100>{130} GB at its peak damage state and after 10 ps. Atoms are colored according to the magnitude of their displacement with arrows highlight final displacement events. GB and surface atoms are grey while all BCC atoms with zero displacement have been removed for clarity.

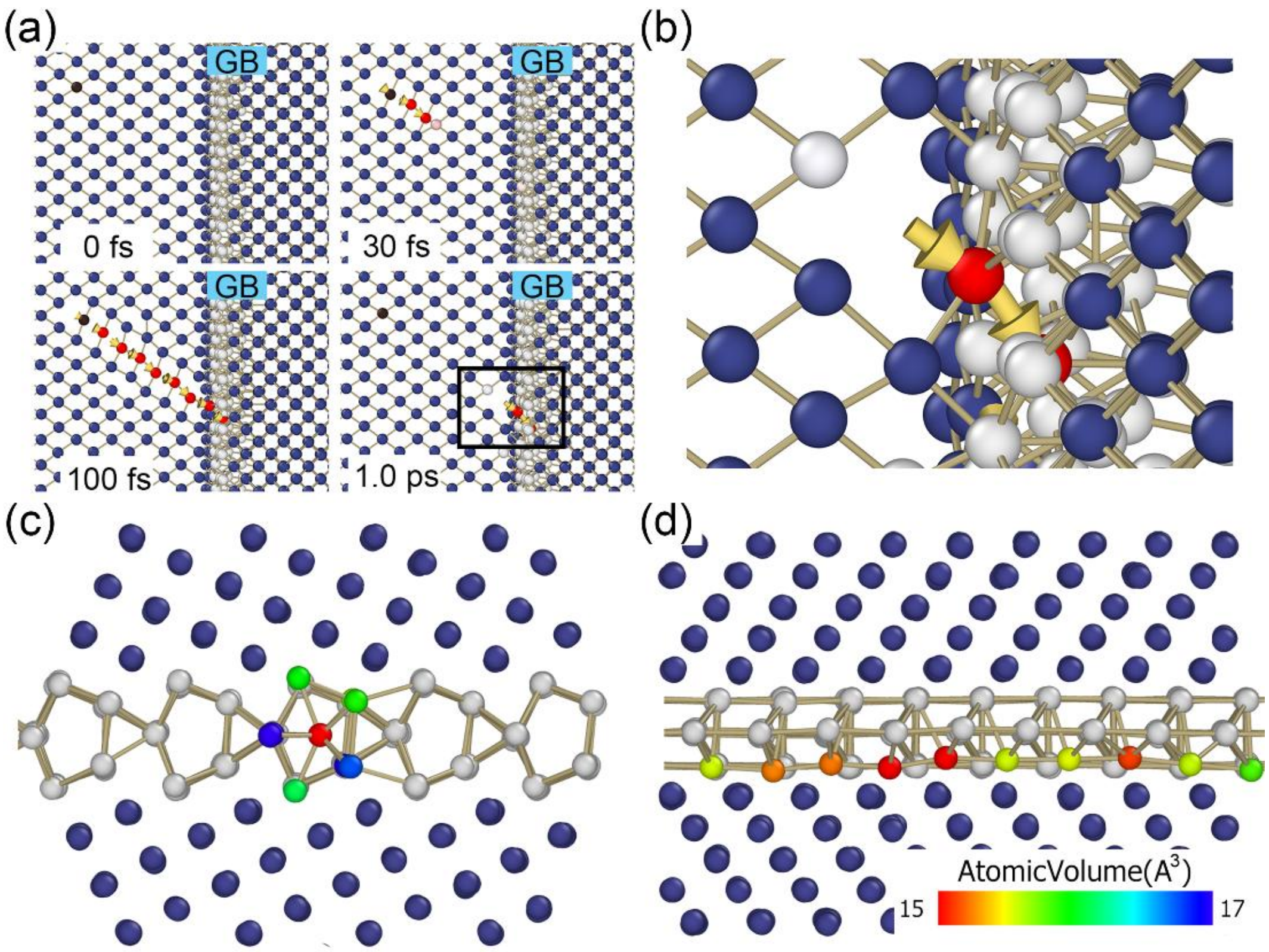

**Figure 10.** (a) Controlled collision sequence with a 60 eV PKA initiated toward a Σ5<100>{130}grain boundary along <111> direction. The collision sequence becomes trapped within the grain boundary plane. (b) Formation of a Frenkel pair near the GB as the collision sequence traverses the GB interface with resulting GB interstitial configurations for (c) the Σ5 GB and (d) the Σ3 GB. Atoms within the interstitials are colored according to their atomic volume while lattice BCC atoms are blue and GB atoms grey.

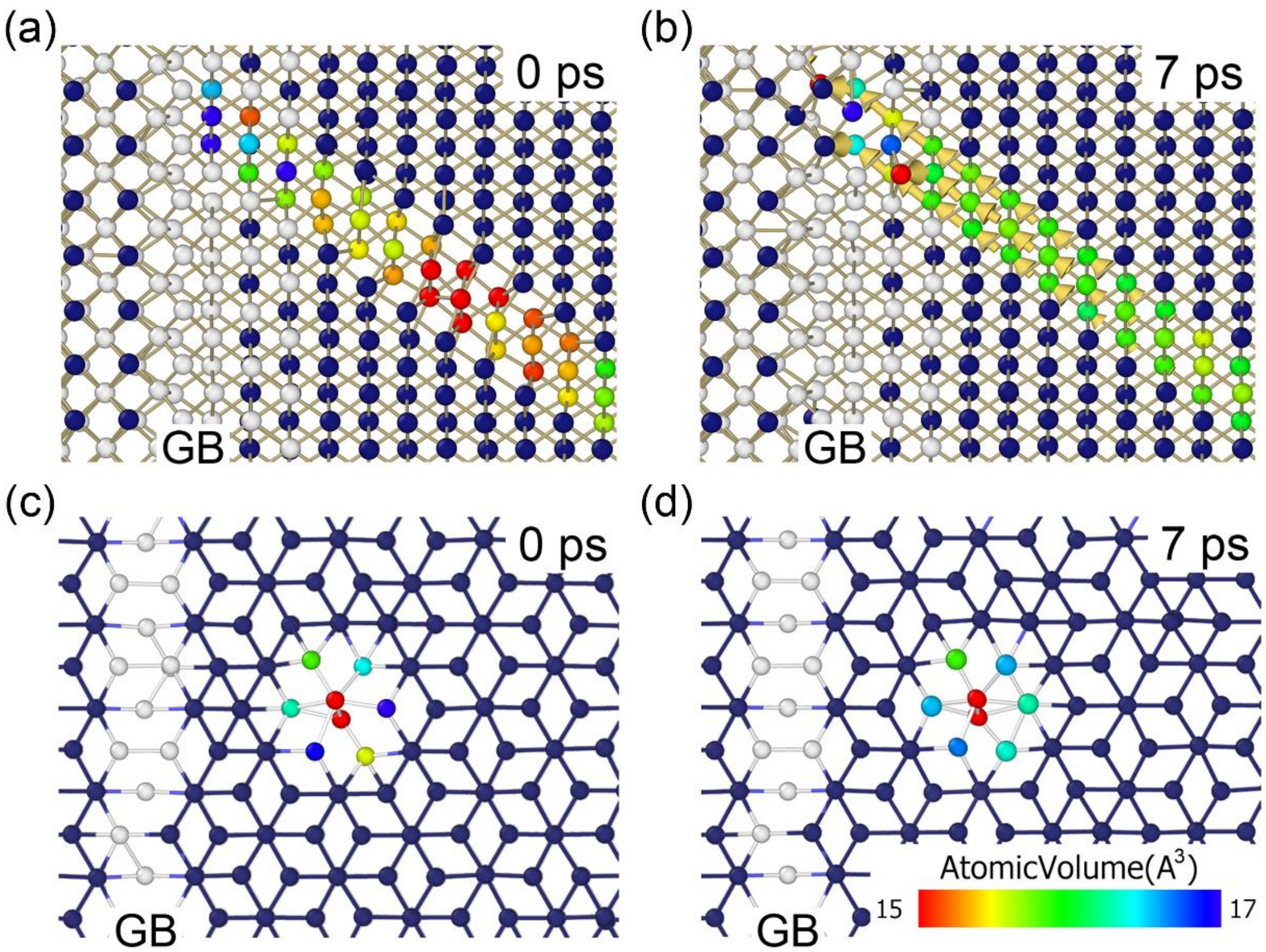

**Figure 11.** A tri-crowdion in the 6 nm nano-bicrystal adjacent to a Σ5<100>{130} GB (a) immediately after its formation and (b) at 7 ps when the crowdion migrated to the GB. A crowdion near a Σ3<110>{112} GB (c) immediately after its formation and (d) 7 ps after where the crowdion has rotated to form a sessile <11ξ> dumbbell configuration. Atoms within the interstitial defects are colored according to their atomic volume while BCC atoms are blue and GB atoms grey.

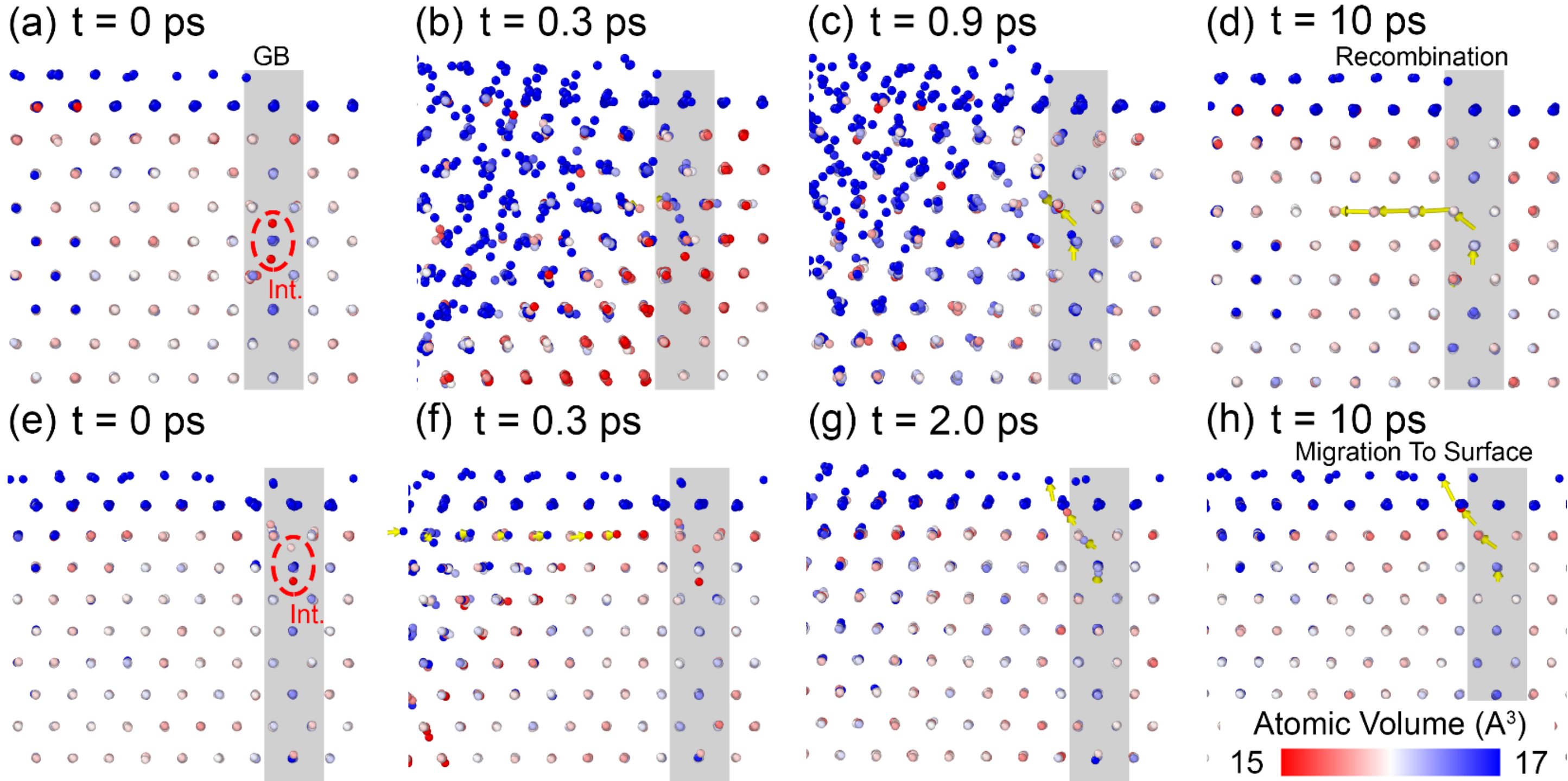

**Figure 12.** Demonstration of the grain boundary interstitial annihilation through (a-d) recombination with a lattice vacancy during a nearby cascade event and (e-h) migration to the surface due to an intrinsic stress gradient. For each mechanism, figure panels are organized from left-to-right by increasing simulation time, and the yellow arrows indicate the interstitial diffusion path during recombination (upper) and surface migration (lower). All atoms are colored according to their atomic volume with initial interstitial positions identified in (a) and (e).

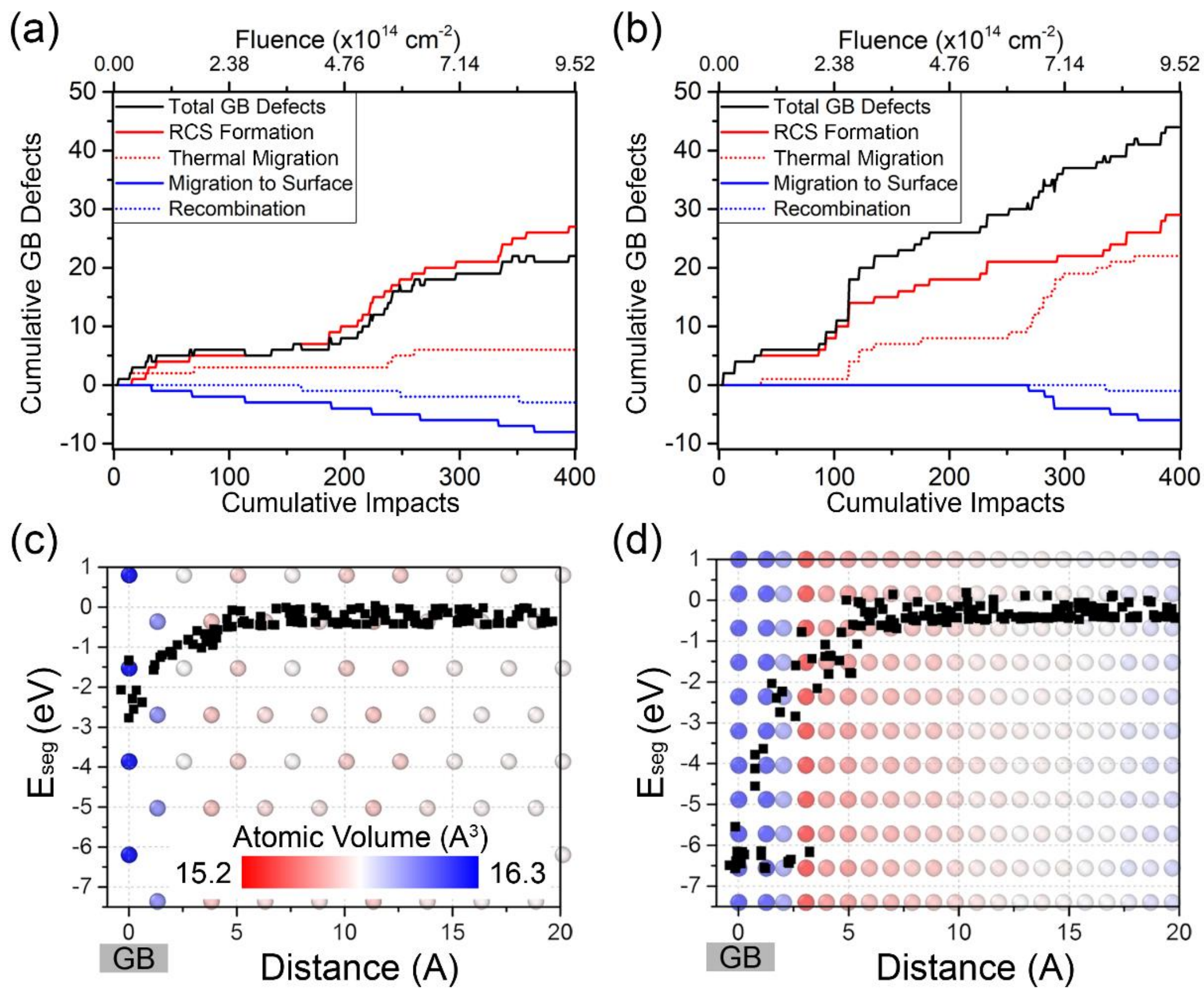

**Figure 13.** (a-b) GB interstitial accumulation trends in the 6 nm nano-bicrystal containing (a) Σ3<110>{112} and (b) Σ5<100>{130} GBs. Recognizing the mechanisms of defect accumulation and annihilation, the cumulative defect trends are delineated into contributions from RCS formation, thermal migration, recombination, and migration to the surface as indicated in the legend. Interstitial segregation energy as a function of distance from the center of the GB plane for (c) the Σ3<110>{112} and (d) the Σ5<100>{130} GBs superimposed over each lattice with atoms colored according to their atomic volume.